\documentclass[journal]{IEEEtran}

\usepackage{graphicx}
\usepackage{amsmath,amssymb} 
\usepackage{color}
 \usepackage{bbm, dsfont}
\usepackage{multirow}
 
\usepackage{longtable}
\usepackage[T1]{fontenc}
\usepackage{epsfig}
\usepackage{authblk}
 
\usepackage{floatrow}
\usepackage{caption}
\usepackage{hyperref}

\usepackage{graphicx}
\usepackage{comment}
\usepackage{amsmath,amssymb} 
\usepackage{color}
\usepackage{subfigure}
\usepackage[]{animate}
\usepackage{animate}
\usepackage{float}
\usepackage{bm}
\usepackage{multirow}
\usepackage{rotating}
\usepackage{wrapfig}
\usepackage{lineno}
\usepackage{epsfig}
\usepackage{nicefrac}       
\usepackage{microtype} 
\usepackage{float}
\usepackage{amsmath}
\usepackage{enumerate}
\usepackage{enumitem} 
\usepackage{soul}
\usepackage{graphicx}
\usepackage{amsmath,amssymb} 
 
\usepackage{color}

\usepackage[noend]{algpseudocode}
\usepackage{multirow}
 
\usepackage{longtable}
\usepackage[T1]{fontenc}
\usepackage{epsfig}

\usepackage{floatrow}
\usepackage{caption}

\usepackage{url}

\usepackage{hyperref}

\usepackage{mathtools}
\usepackage{resizegather}
\usepackage{dsfont}
\usepackage{caption}
\usepackage{multirow}

\usepackage{amsthm}
\theoremstyle{plain}

\usepackage[section]{placeins}

\newfloatcommand{capbtabbox}{table}[][\FBwidth]
\usepackage[table]{xcolor}
\definecolor{city_color_0}{rgb}{0.0,0.0,0.0}
\definecolor{city_color_1}{rgb}{0.5020,0.2510,0.5020}
\definecolor{city_color_2}{rgb}{0.9569,0.1373,0.9098}
\definecolor{city_color_3}{rgb}{0.2745,0.2745,0.2745}
\definecolor{city_color_4}{rgb}{0.4000,0.4000,0.6118}
\definecolor{city_color_5}{rgb}{0.7451,0.6000,0.6000}
\definecolor{city_color_6}{rgb}{0.6000,0.6000,0.6000}
\definecolor{city_color_7}{rgb}{0.9804,0.6667,0.1176}
\definecolor{city_color_8}{rgb}{0.8627,0.8627,0.0000}
\definecolor{city_color_9}{rgb}{0.4196,0.5569,0.1373}
\definecolor{city_color_10}{rgb}{0.5961,0.9843,0.5961}
\definecolor{city_color_11}{rgb}{0.2745,0.5098,0.7059}
\definecolor{city_color_12}{rgb}{0.8627,0.0784,0.2353}
\definecolor{city_color_13}{rgb}{1.0000,0.0000,0.0000}
\definecolor{city_color_14}{rgb}{0.0000,0.0000,0.5569}
\definecolor{city_color_15}{rgb}{0.0000,0.0000,0.2745}
\definecolor{city_color_16}{rgb}{0.0000,0.2353,0.3922}
\definecolor{city_color_17}{rgb}{0.0000,0.3137,0.3922}
\definecolor{city_color_18}{rgb}{0.0000,0.0000,0.9020}
\definecolor{city_color_19}{rgb}{0.4667,0.0431,0.1255}

\usepackage{epsfig}
\usepackage{amsmath,amssymb} 
\usepackage{multirow}

\makeatletter
\def\ps@IEEEtitlepagestyle{
  \def\@oddfoot{\mycopyrightnotice}
  \def\@evenfoot{}
}
\def\mycopyrightnotice{
  {\footnotesize
  \begin{minipage}{\textwidth}
  \centering
  978-1-6654-2113-3/21/\$31.00 \copyright2021 IEEE
  \end{minipage}
  }
}

\begin{document}
 
\title{Posterior Estimation for Dynamic PET imaging using Conditional Variational Inference}

\author{Xiaofeng~Liu,
        Thibault~Marin,
        Amal~Tiss,
        Jonghye~Woo,
        Georges~El Fakhri,
        and~Jinsong~Ouyang



\thanks{X. Liu, T. Marin, A. Tiss, J. Woo, G. El Fakhri, and J. Ouyang$^*$ are with the Gordon Center for Medical Imaging, Massachusetts
General Hospital and Harvard Medical School, Boston, MA 02114 USA.($^*$corresponding author: ouyang.jinsong@mgh.harvard.edu)}
}\vspace{-25pt}

\maketitle

\begin{abstract}
This work aims efficiently estimating the posterior distribution of kinetic parameters for dynamic positron emission tomography (PET) imaging given a measurement of time of activity curve. Considering the inherent information loss from parametric imaging to measurement space with the forward kinetic model, the inverse mapping is ambiguous. The conventional (but expensive) solution can be the Markov Chain Monte Carlo (MCMC) sampling, which is known to produce unbiased asymptotical estimation. We propose a deep-learning-based framework for efficient posterior estimation. Specifically, we counteract the information loss in the forward process by introducing latent variables. Then, we use a conditional variational autoencoder (CVAE) and optimize its evidence lower bound. The well-trained decoder is able to infer the posterior with a given measurement and the sampled latent variables following a simple multivariate Gaussian distribution. We validate our CVAE-based method using unbiased MCMC as the reference for low-dimensional data (a single brain region) with the simplified reference tissue model.


\end{abstract}


\IEEEpeerreviewmaketitle

\vspace{-5pt}
\section{Introduction}

For many physical imaging systems, a common problem is to determine hidden system parameters $\bm{x}$ from a measurement $\bm{y}$. For example, in kinetic modeling of tau PET studies, our goal can be inferring the posterior distribution of $p(\bm{x}|\bm{y})$ w.r.t. distribution volume ratio (DVR) for a given measurement of time-activity curve (TAC) in a target brain region. 

Usually, the physical model describing how measurable quantities $\bm{y}$ arise from the hidden parameters $\bm{x}$ is well-defined by a forward process. For example (used in this work), the forward process can be,  $\bm{y}=f(\bm{x})+\varepsilon$, where $\bm{x}=\{DVR,k_2,R_1\}$ in a target brain region ($k_2$ is the rate constant from free to plasma compartment, $R_1$ is the ratio of rate constants for transform from plasma to free compartment), $\bm{y}$ is the measured TAC in the region, $f$ represents the kinetic modeling using simplified reference tissue model (SRTM) \cite{lammertsma1996simplified} in PET, and noise $\varepsilon$ follows a Gaussian distribution. As a result, the likelihood, $p(\bm{y}|\bm{x})$, is defined. If we assume a prior, $p(\bm{x})$, based on our knowledge before the measurement. The posterior distribution is determined as $p(\bm{x}|\bm{y}) \propto p(\bm{y}|\bm{x})p(\bm(x))$.  The conventional approach to sample the posterior distribution is to follow a rejection sampling scheme with Markov Chain Monte Carlo (MCMC) \cite{andrieu2003introduction}. MCMC is known to produce unbiased asymptotical estimation \cite{andrieu2003introduction}, while it is impractical for high-dimensional data due to its high computational cost. In this work, we propose a deep-learning-based approach to sample posterior distribution.



\vspace{-5pt}
\section{Methodology}

In this work, we resort to the deep variational Bayes for the estimation of the posterior distribution. Specifically, we propose to use a conditional variational autoencoder (CVAE) framework and optimize its evidence lower bound (ELBO) for accurate posterior estimation. To counteract the inherent information loss of the forward process, we introduce latent variables $\bm{z}$, which capture the loss of information in the forward process. Thus, our approach explicitly learns to associate hidden parameters $\bm{x}$ with the unique pairs [$\bm{y}$, $\bm{z}$] of measurements and latent variables.

\begin{figure}[!t]
\centering
\includegraphics[width=8cm]{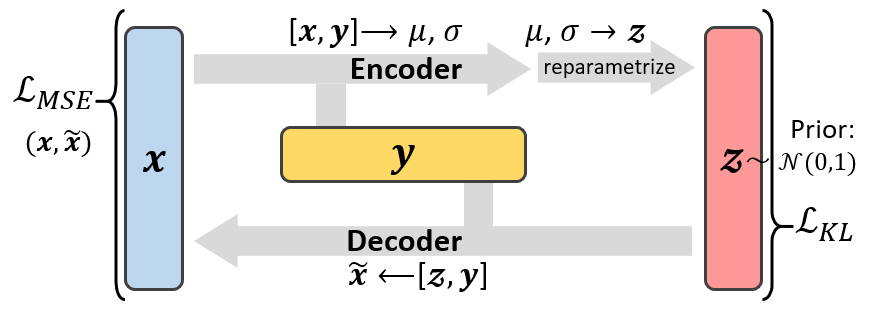}\vspace{-2pt}
\caption{The posterior estimation framework using CVAE. The network is optimized with loss function $\mathcal{L}=\mathcal{L}_{MSE}+\mathcal{L}_{KL}$.\vspace{-10pt}}\label{fig:1}\vspace{-10pt}
\end{figure}

With the paired data in $\{\bm{x}_i,\bm{y}_i\}_{i=1}^M$, we reconstruct $\bm{x}$ to ensure $\bm{y}$ and $\bm{z}$ have complementary information to recover $\bm{x}$. Additionally, we make sure that the density $p(\bm{z})$ of the latent variables is shaped as a multivariate Gaussian distribution. Thus, the CVAE represents the desired posterior $p(\bm{x}|\bm{y})$ by a deterministic trained decoder, i.e., $\bm{x}=$ Decoder$(\bm{y},\bm{z})$, that transforms the known distribution $p(\bm{z})$ to $\bm{x}$-space, conditional on $\bm{y}$. The general idea is illustrated in Fig.~\ref{fig:1}.

Based on the well-defined forward process, our CVAE framework aims to efficiently infer the posterior $p(\bm{x}|\bm{y})$ for a given measurement $\bm{y}$ and prior $p(\bm{x})$. The encoder $[\bm{x},\bm{y}]\rightarrow {\bm{z}}$ encodes the latent variation $\bm{z}$ with $\bm{x}$ and $\bm{y}$. We would expect $\bm{z}$ to capture the information lost in the forward process. This is implicitly achieved by enforcing a information bottleneck \cite{alemi2016deep} of the small size $\bm{z}$ to have sufficient content to recover $\bm{x}$. The decoder $[\bm{z},\bm{y}]\rightarrow \tilde{\bm{x}}$ reconstructs the parameters $\bm{x}$ with $\bm{z}$ and $\bm{y}$. We empirically use three fully connected layers for both encoder and decoder. Similar to original VAE \cite{kingma2013auto}, the encoder has two output vectors, $i.e.,$ $\mu$ and $\sigma$. We then utilize the reparametric trick $\bm{z}=\mu+\sigma\odot\epsilon$, where $\epsilon\in \mathcal{N}(0,I)$. The posterior distribution of $\bm{z}$ is $q(\bm{z}|\bm{x},\bm{y})\sim \mathcal{N}(\bm{z};\mu,\sigma^2)$.

\begin{figure*}[!t]
\centering
\includegraphics[width=17cm]{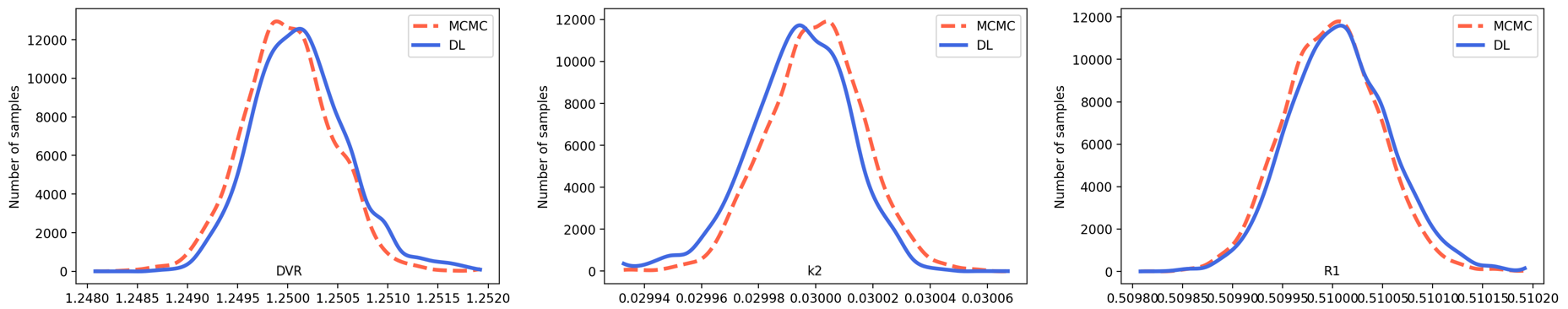}\vspace{-10pt}
\caption{Comparison of the MCMC and our deep learning (DL) method. We show the smoothed curves of 25 bins histograms. We set $DVR=1.25$, $k_2=0.03$ and $R_1=0.51$ to calculate $\bm{y}^*$ with SRTM and sample 90,000 $\bm{z}$ following a Gaussian prior. \vspace{-10pt}} \label{fig:3}
\end{figure*}

\begin{figure}[!t]
\centering
\includegraphics[width=8cm]{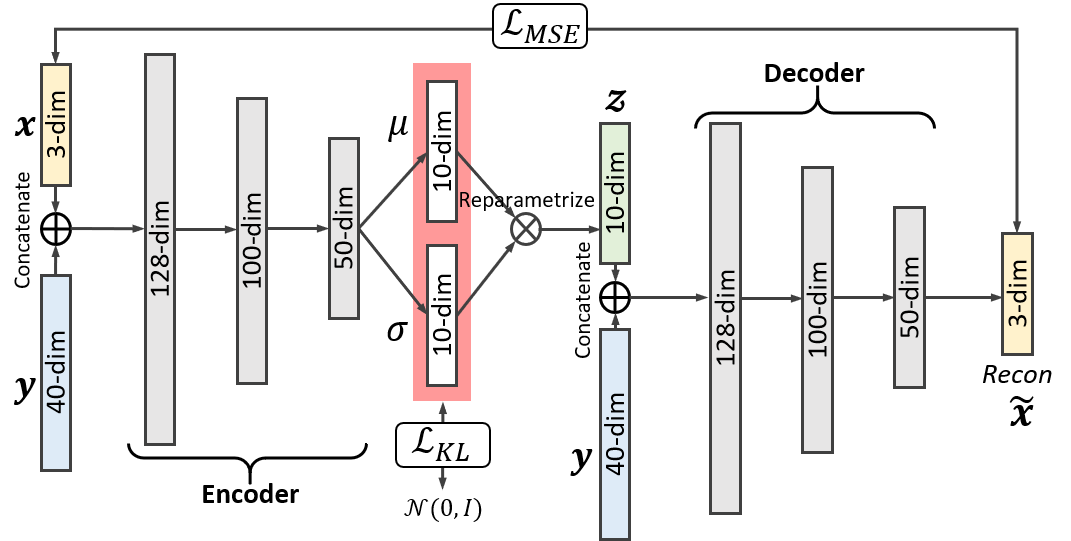}\vspace{-10pt}
\caption{Detailed CVAE-based posterior estimation framework for a single voxel. Only decoder is used for testing.\vspace{-10pt}} \label{fig:2}
\end{figure}

In training, the ELBO \cite{kingma2013auto} of our CVAE includes the Kullback-Leibler (KL) divergence and the reconstruction loss. In practice, the KL-divergence can be computed as\vspace{-10pt} {\begin{equation}
\begin{aligned} 
   \mathcal{L}_{KL}(\bm{z};\mu,\sigma)=\frac{1}{2}\sum^{M_z}_{j=1}(1+{\rm log}(\sigma_j^2)-\mu_j^2-\sigma_j^2), \label{eq:m1} \vspace{-10pt}
\end{aligned}\end{equation}}where $M_z=10$ is the dimension of the latent code $\bm{z}$. For the reconstruction error, we simply adopt the pixel-wise mean square error (MSE). Let $\tilde{\bm{x}}$ be the reconstructed $\bm{x}$, their MSE loss can be formulated as:\vspace{-10pt}{\begin{equation}
\begin{aligned}
    \mathcal{L}_{MSE}(\bm{x},\tilde{\bm{x}})=\frac{1}{2}\sum^{M_{x}}_{j=1}||\bm{x}_{j}-\tilde{\bm{x}}_{j}||^2_2,\label{eq:m2}\vspace{-10pt}
\end{aligned}\end{equation}}where ${M_{x}}=3$ indicates the dimension of $\bm{x}$ or $\tilde{\bm{x}}$.

In testing, only the decoder is used. Given an observation $\bm{y}^*$, we sample $\bm{z}\sim \mathcal{N}(0,I)$ and concatenate each of them with $\bm{y}^*$. If ${\bm{x}}\sim p(\bm{x}|\bm{y}^*)$ and $\bm{y}=\bm{y}^*$, we have following two claims: 1) $\tilde{\bm{x}}\sim p(\bm{x}|\bm{y}^*)$ because $\mathcal{L}_{MSE}=0$; 2) the corresponding conditional probability $q(\bm{z}|\bm{x},\bm{y}^*)$ is equivalent to $\mathcal{N}(0,I)$ because $\mathcal{L}_{KL}=0$. Since the distribution of $\tilde{\bm{x}}$ is solely determined by $\bm{z}$ distribution and $\bm{y}$ input, the above two claims guarantee that the decoder with inputs $\bm{z}\sim \mathcal{N}(0,I)$ and $\bm{y}=\bm{y}^*$ outputs posterior distribution $p(\bm{x}|\bm{y}^*)$.






\section{Experiments and results}

The approach described above can be applied to many medial imaging problems. In this work, we estimate the posterior distribution of DVR in a single brain region given a measurement of TAC in the region using both our CVAE-based approach and MCMC. The unbiased distribution estimated by MCMC is the reference used to validate our approach.  The forward process is defined by a kinetic model and a noise model. The kinetic model is SRTM written as:
\begin{eqnarray}\vspace{-10pt}
\frac{d C_T(t)}{dt} = R_1 \frac{d C_R(t)}{dt} + k_2C_R(t) - \frac{k_2}{DVR}C_T(t),\label{eq1}\vspace{-10pt}
\end{eqnarray}
where $C_T(t)$ and $C_R(t)$ are the activity concentrations in the target and reference regions, respectively. The noise model is defined as:  $y_n=\int_{t_{n-1}}^{t_n} C_T(t) \mathrm{d}t + \epsilon_n$, where $\frac{\epsilon_n}{\sigma\sqrt{\Delta t_n/T}}\sim \mathcal{N}(0,1)$ and $T=\sum_{n=1}^N \Delta t_n$, $\Delta t_n={t_n}-{t_{n-1}}$, $\sigma$ is the standard deviation. 




With enough samples, the MCMC algorithm would produce the unbiased posterior samples asymptotically for Bayesian inference \cite{chib1995understanding}.  Without loss of generality, we simply set the number of time frame $N=40$, $\Delta t_n=0.5$ min in the first ten frames and $\Delta t_n=1$ min in the later 30 frames, and $\sigma\sim$ Gamma(1,1) in our forward model. We generate $M=10,000$ samples for training. Specifically, we sample $DVR\sim\mathcal{N}(1,1)$, $k_2\sim\mathcal{N}(0.05,0.01)$, $R_1\sim\mathcal{N}(0.5,0.1)$, and only select the positive values.
 
Metropolis-Hastings sampler (MHS) \cite{chib1995understanding}, a variant of MCMC, was used to sample from the ground truth posterior. In testing, we performed 100,000 iterations of random walk Metropolis sampling for a generated measurement using SRTM with 10,000 burn-in steps. With the SRTM in a single voxel, PyMC\footnote{\url{https://pymc-devs.github.io/pymc/}} takes about 10 mins to infer 90,000 samples. In testing, our learning-based CVAE can infer 90,000 samples of a given $\bm{y}^*$ and the sampled $\bm{z}$ within 30s. We note that MCMC takes a similar process time for every $\bm{y}^*$. In contrast, the trained CVAE can make an efficient posterior estimation for different $\bm{y}^*$. 

The detailed network implementation using our approach is shown in Fig.~\ref{fig:2}. We construct the paired dataset $\{\bm{x}_i,\bm{y}_i\}_{i=1}^M$ by randomly sampling $\bm{x}_i$ and $\sigma$ following a Gaussian and Gamma distribution, respectively. Then, we generate its corresponding $\bm{y}_i$ with the forward kinetic model. 

 The comparison of the posterior estimation results is shown in Fig.~\ref{fig:3}. The posterior distributions using MCMC and our proposed approach are in good agreement. 

\section{Conclusions}
We have proposed a CVAE framework for efficient posterior estimation of dynamic PET imaging. The preliminary results for sampling DVR posterior distribution show good agreement between unbiased MCMC and our proposed approach. Our approach has much faster inference than MCMC. In the future, we will apply it to high-dimensional imaging data.

\bibliographystyle{ieee}
\bibliography{egbib2}

\end{document}